\documentstyle[aps,prl,epsfig]{revtex}

\def\f{\phi}
\def\L{\Lambda}
\def\l{\lambda}
\def\t{\tau}

\def\u{\underline}

\def\F{\Phi}
\def\a{\alpha}
\def\b{\beta}
\def\s{\sigma}
\def\AH{\ ^{(\a)}H  }
\def\AX{\ ^{(\a)}X  }
\def\OX{\ ^{(0)}X  }
\def\Ah{\ ^{(\a)}H  }
\def\Oh{\ ^{(0)}H  }
\def\APX{\ ^{(\a)}{\mathcal{X}}  }
\def\OPX{\ ^{(0)}{\mathcal{X}}  }

\begin{document}

\title{A systematic study of the radion in the compact Randall-Sundrum model}

\author{ Ashok Das and Alexander Mitov}

\address{Department of Physics and Astronomy,
University of Rochester, NY 14627-0171, USA}

\maketitle \vskip .5cm

\begin{abstract}

We systematically study the question of identification and consistent
inclusion of the radion, within the Lagrangian approach, in a two
brane  Randall-Sundrum model. Exploiting the symmetry
properties of the theory, we show how the radion can be identified
unambiguously and give the action to all orders in the radion field
and the metric. Using the background field method, we expand the
theory to quadratic orders in the fields. We show that the most
general classical solutions, for the induced metric on the branes in
the case of a constant radion and a factorizable $4D$ metric,
correspond to  Einstein spaces. We discuss extensively
the diagonalization of the quadratic action. Furthermore, we obtain
the 4-dimensional effective theory from this and study the question of
the spectrum as well as the couplings in these theories.

\end{abstract}
\vspace{.7in}

\section{Introduction}

The idea that we may live in a world of more than four space-time
dimensions is certainly not new \cite{GSW}, but in recent years it has
received a
lot of attention because of the observation that the extra compact
dimensions need not all be of the order of the Planck length scale.
A large extra dimension of the order of a $(TeV)^{-1}$ \cite{Dvali}
provides  an alternate
solution to the hierarchy problem (the desert scenario), namely, it
explains in a natural manner the large ratio for the Planck scale to
the electro-weak scale. There are basically two interesting approaches
to solving this problem. The first, due to Arkani-Hamed, Dimopoulos and
Dvali  \cite{Dvali}, is based on the  standard Kaluza-Klein
approach, where the higher dimensional space is a direct product of
the $4$ dimensional space-time and a compact manifold. The second, due
to Randall and Sundrum (RS) \cite{RSfirst,RSsecond}, uses an alternate
and  interesting scenario,
where the higher dimensional space-time is not factorizable.

The simplest RS model can be envisaged as follows. Let us consider a
five  dimensional
space-time manifold, with the extra coordinate taking values on an
orbifold $S^1/Z_2$, with two fixed points, which are
chosen to be at, say, $0$ and $\pi$. At each of the two fixed points of the
orbifold, there is a singular source  - a $3$-brane - which carries a
constant, nonzero tension. In addition, a negative five dimensional
cosmological constant is also assumed to be present in the bulk, which
leads to a five-dimensional metric that is a slice of the AdS space
\cite{RSfirst}. Since the warp factor turns out to be an exponential
function depending on the extra coordinate, the ratio of the values of
the  warp factor at the two
branes, for such a metric, naturally leads to the ratio of the
Planck to $TeV$ scales without the need for an extremely large extra
dimension.  We can think of our physical world as
described by one of the two branes, conventionally chosen to be the one at
$\pi$. In the original RS model, the values
of the five dimensional cosmological term and the tensions of the two
branes are chosen in such a way, that the effective
four dimensional cosmological term is zero and 
the induced metric is just the Minkowski metric. In this model, the
physical distance between the two branes is an arbitrary constant
$r_c$, whose value cannot be determined from the classical analysis. The
stabilization of the size of the extra dimension requires additional
mechanisms \cite{GW}. Even when the five dimensional
cosmological constant and the brane tensions are not fine tuned to
cancel (as is the case in the original RS model),
it was shown in \cite{dSAdS} that the model admits solutions with
induced  metrics of dS or AdS type. In such a case, the distance between
the two branes can be determined in terms of the parameters of the model.

The classical solutions of the RS model, as discussed above, correspond
only to 
the vacuum sector of the theory when there is no matter present.
One can, of course, add matter to the theory, which is assumed to be
located on  the physical brane. In principle, one can also add bulk
matter. In addition, one can also analyze the fluctuations of the five
dimensional gravity around the vacuum solution, as well as the
fluctuation of the distance between the branes. In this
paper, we will not consider the inclusion of matter, and will
investigate only the contributions arising from the fluctuations of the
metric and the geometry. To fix the terminology, let us note that the
four  dimensional scalar field, which describes the distance between
the two branes, is called the radion. The vacuum expectation value of
this  field is expected to determine the constant $r_c$, although we
will not worry about the mechanism that leads to this vacuum
expectation value. The radion was first introduced in
\cite{RSfirst,RSsecond}, where fluctuations of gravity and the
radion were simply added to the vacuum values
of the metric and $r_c$ respectively. Although this is  quite natural
and is standard in quantum field theory, it was
later pointed out in \cite{Rubakov} that the ansatz of \cite{RSfirst,RSsecond}
does not  solve the
linearized Einstein equations. Instead, the authors of \cite{Rubakov}
used an alternate form of the metric, that is correct to linear
order in the radion fluctuations. 

The inclusion of the radion field was of  great importance
in the construction of the supersymmetric version of the RS model
\cite{Bagger,Luty}, where the bosonic sector of the
theory, in addition to other fields, contains the radion and the graviton (but
not its KK tower) to all orders in these fields. The ansatz 
used in \cite{Bagger} coincides to linear order with that of \cite{Rubakov},
while that in \cite{Luty} corresponds to \cite{RSfirst,RSsecond}. The two
papers  contain
similar results, which are valid up to second order in the  space-time
derivatives. The inclusion of the radion in the
case of  dS and AdS branes has also been discussed in \cite{Fox},
using an ansatz of the form of \cite{RSfirst,RSsecond}.

We believe that, at present, we do not have a complete understanding
of the radion in the RS model and it is the proper identification and
the consistent inclusion of the radion field in an arbitrary two brane
model which is the purpose of this paper. We 
derive the complete five dimensional Lagrangian for the RS
model, including the radion and the graviton (with its KK tower), and
study systematically the question of the spectrum of the theory by
restricting the Lagrangian up
to second order in these fields. We follow the
geometrical approach of \cite{Rubakov}, where the authors
identified the radion field from an analysis of the symmetries of the
Einstein equations in the bulk, and derived the form of the five dimensional
metric up to linear order in the radion. The work of \cite{Rubakov}
emphasizes  the
importance of the junction conditions which, as we will see, are
of utmost importance in our investigation as well and are quite relevant
for understanding the mixing between the radion and the graviton
and its KK tower. However, in our opinion, there are some features in
the  analysis of \cite{Rubakov}
that remain obscure and we study this problem systematically from the
conventional Lagrangian approach.
We formulate the RS model as the model of five dimensional
space-time with two 3-branes at the fixed points of an orbifold, whose
action is manifestly invariant under arbitrary five dimensional
transformations, and we carefully elucidate the symmetry
properties of the model. Such an approach makes the identification
of the radion field unambiguous and free of assumptions. Subsequently,
employing the background field method, we are able to make a very
general and systematic investigation of the possible vacuum solutions
in the  RS model as well as give a  satisfactory $5D$ Lagrangian
description (that holds for any vacuum solution) of the radion
and the graviton and its KK tower, up to second order in the
fields, without mixing between them.

The paper is organized as follows. In section {\bf 2}, we describe the
model, both in an interval as well as on the orbifold. Taking
advantage of the symmetries of the theory, we parameterize the metric
in a manner that makes the identification of the radion natural and is
convenient for our subsequent discussions. The Lagrangian density of
the theory is then expressed completely in terms of this
parameterization to all orders in the field variables. In section {\bf 3},
we use the background field method to expand the action as well as the
junction conditions to quadratic order in the field variables. The
classical equations are solved in a unified manner and we show that
a factorizable background metric, in general, defines an Einstein
space multiplied by a warp factor. We then discuss
the question of diagonalization of the quadratic action as well as
the boundary conditions in a systematic manner. In section {\bf 4}, we
rewrite this action as an effective action in 4-dimensions and discuss
further the properties of this diagonalized theory. In section {\bf
5}, we present a brief summary of our results. In appendix {\bf A}, we
describe some relations that are useful in the background field
expansion of the theory while in appendix {\bf B}, we discuss the
Kaluza-Klein decomposition of the metric as well as some of the
properties of the basis functions.

\section{The Model}

Let us consider a  $5$-dimensional space-time manifold with signature
$(-,+,+,+,+)$, which is parameterized by the usual $4$-dimensional space-time
coordinates as well as a fifth coordinate that is bounded by the
locations of two $4$-dimensional hypersurfaces. The hypersurfaces can
be specified by equations of the forms 
\begin{equation}\label{ffi}
f_L(x,z)=0,\qquad f_R(x,z)=0
\end{equation}
The functions $f_{L,R}$ are, in general, arbitrary except for the
condition that everywhere on the boundary hypersurfaces,
the normal fields, given by $n=df_{L,R}(x,z) \vert_{f=0}$, are
space-like.  The locations of the two hypersurfaces can be determined
by inverting (\ref{ffi}), namely,
\begin{equation}\label{ffii}
z_L=\f_L(x)\ ,\qquad z_R=\f_R(x)
\end{equation}
where $\f_{L,R}(x)$ are not necessarily small.
Therefore,
$  z\in I_z:=[\f_L(x), \f_R(x)].  $
From (\ref{ffii}) it is clear that under arbitrary $4D$
transformations, $z'=z\ ,\ x'=x'(x)$, the functions $\f_{L,R}$ 
transform as $4D$ scalars.

On this $5D$ space-time manifold , one can define a theory which
is invariant under arbitrary $5D$ transformations. The
action has the form
\begin{equation}\label{action}
S\ =\ S_{\rm bulk}\ +\ S_L\ +\ S_R
\end{equation}
where
\begin{eqnarray}
S_{\rm bulk}\ &\propto &\ \int_{{\rm bulk}} d^5V \sqrt{-G}\left( -R^{(5)}-\L +\dots\right)\label{bulk}\\
S_{i}\ &\propto &\ \int_{f_i=0} d^4 V \sqrt{-g^i}
\left(-{\cal V}_{i} +\dots\right),\qquad i={L,R} \label{branes}
\end{eqnarray}
Here, $G_{MN}$ and $G$ are the $5D$ metric and its
determinant respectively, while $g_{mn}^{i}$ is the $4D$ induced
metric on the respective boundary hypersurface. In this paper,
we use the convention that capital roman letters denote $5D$
indices, while lower case roman letters represent $4D$ indices. We
also use
the notations and definitions of \cite{Weinberg} throughout. $\L$ is a $5D$
cosmological constant, ${\cal V}_{i}$ are the tensions on each of the
branes and the dots denote possible terms representing $5D$ as well as
$4D$ matter, which we ignore in the present study.

To obtain the equations of motion, we need to vary the action
(\ref{action}). As long as the positions of the boundaries are not
specified (i.e. $f_{L,R}$ are arbitrary), it is not enough to
consider the variation of the metric alone. We must also extremize the
action  with respect to the ``volume" where the theory is defined. To do so,
it will be more convenient to make use of the $5D$ coordinate
invariance of  the model and choose a special coordinate 
system, where the volume does not depend on
dynamical quantities such as $f_{L,R}$ and, consequently, is not
subject to variations.

Such a transformation can be done in two steps. First, we adopt
Gaussian Normal (GN) coordinates with respect to the brane $\f_L$. In
this  case, we
have $\f_L= {\rm constant}$, which we choose to be zero, namely,
$\f_L=0$. In this coordinate system, the second boundary is located at
$\f_R=\tilde{\f}(x)$ and the $5D$ metric takes the form, $G_{m5}=0\
,\ G_{55}=1$. Clearly, the function $\tilde{\f}(x)$ represents the
physical distance between the two boundaries, and can be related to
the radion field.
Next, we rescale the fifth coordinate $z=\tilde{\f}(x)
t/\pi\ ,\ t\in I_t:=[0,\pi]$ ($t$ does not correspond to time which is
denoted by $x^{0}$). Introducing a new function
$\f(x):=\tilde{\f}(x)/\pi$, we have the two boundaries at the
fixed end points $t=0$ and $t=\pi$ respectively and the $5D$ volume
takes the form: $M^4\times I_t$. (A more general
transformation of this type has already been discussed in \cite{PRZ}.)
Note that neither the
above transformations nor the field $\f(x)$, which we 
call radion, have to be small. It is worth noting here that $\f(x)$ is
a four dimensional scalar field as the radion should be and it has no
dependence on the extra coordinate, a reflection of the fact that
there is no Kaluza-Klein tower for the radion. 

In these coordinates, the $5D$ metric takes the form
\begin{eqnarray}\label{metric}
 G_{MN}= \left(
         \begin{array}{cc}
              g_{mn}(x,t)          &  N_n        \\
                                   &             \\
              N_m                  &   \f^2(x)   \\
          \end{array} \right)
\end{eqnarray}
and the inverse metric is determined to be
\begin{eqnarray}\label{inverse}
 G^{MN}= \left(
         \begin{array}{ccc}
              g^{mn}(x,t)+{N^m N^n\over N} &              &  -{N^n\over N}  \\
                                           &     ~~~~~    &                 \\
              -{N^m\over N}                &              &   {1\over N}    \\
          \end{array} \right)
\end{eqnarray}
where the raising and the lowering of the $4D$ indices, $m,n,\dots$,
is done with the metric $g_{mn}(x,t)$ satisfying
$g_{mn}(x,t)g^{nk}(x,t)=\delta_m^k$. Furthermore, $N_m(x,t)= {t\over
2}(\f^2(x))_{,m}$ and $N(x,t)=\f^2(x)-N_m(x,t)N^m(x,t)$ where a comma
denotes a derivative. We see that
the metric in these coordinates has a  form similar to that of
Arnowitt-Deser-Misner \cite{ADM}. Let us note that, in these
coordinates,  the induced
metric on each of the boundaries is given by
\begin{equation}\label{indmetric}
g_{mn}^{i}(x)\ =\ g_{mn}(x,t)\vert_{t=t_i}.
\end{equation}
Furthermore, it is easy to show that 
\begin{equation}\label{det}
\det\left(G_{MN}(x,t)\right)\ =\ N\ \det\left( g_{mn}(x,t)\right).
\end{equation}

This defines the theory on an interval $t\in I_{t}=[0,\pi]$.
Let us next replace the interval, $I_t$, by the orbifold $S^1/Z_2$.
In order to do that, we extend the variable $0\leq t \leq\pi$ to the
interval $[-\pi,\pi]$, imposing the additional symmetry
$t\to -t$. As a result, we need to replace 
$t$ by $\t=|t|$ in the metric (\ref{metric}) so that $g_{mn}(x,\t)$ is a
symmetric and non-degenerate, but otherwise arbitrary tensor and,
\begin{eqnarray}
N_m(x,\t) &=&\ {\t\over 2}(\f^2(x))_{,m} \nonumber \\
N(x,\t) \ &=&\ \f^2(x)-N_m(x,\t)N^m(x,\t)   \label{N}
\end{eqnarray}
Throughout the paper, we will denote by a dot (e.g. $\dot{a}$)
differentiation with respect to $\t$, while a prime (e.g. $a'$)
denotes differentiation with respect to $t$. We denote the $4D$ partial
derivatives by a comma (e.g. $a_{,m}$), while a covariant
derivative is represented by a semicolon.

The action (\ref{action}), on the orbifold, can be written as
\begin{eqnarray}
S\ & = & \ S_{\rm bulk} + \sum_{i=0,\pi} S_{i}\nonumber\\
S_{\rm bulk}\ &=&\ \int d^4x\int_{-\pi}^\pi dt\ \sqrt{-G}\left(
-2M^3R^{(5)}-\L \right)\label{bulki}\nonumber\\
S_{i}\ &=&\ \int d^4 x \int_{-\pi}^\pi dt\ \sqrt{-g^{i}}
\left(-{\cal V}_{i} \right)\delta(t-t_i),\qquad t_i={0,\pi}
\label{branesi}
\end{eqnarray}
where $R^{(5)}$ is the Ricci scalar constructed from
the $5D$ metric $G_{MN}$ given in (\ref{metric}) and generalized to the
orbifold, as discussed above. In studying the solutions of this theory, 
the standard approach would be to solve the Einstein
equations following from the action (\ref{branesi}). However, we note
that, since the $5D$ metric
(\ref{metric}) is already decomposed into a 
$4+1$ form because of the special coordinate choice we have made, it
is natural to recast the action first in terms
of the  $4D$ metric, $g_{mn}(x,\t)$, and the radion field, $\f(x)$,
using this parameterization. After some
lengthy algebra, we obtain
\begin{equation}\label{actiongen}
S\ =\ -2M^3\int d^4x\int_{-\pi}^\pi dt \sqrt{-g}\left({\mathcal{L}}
+ \sum_{i=0,\pi}{\cal L}_{i}\right)
\end{equation}
where
\begin{eqnarray}
{\mathcal{L}}\ &=&\ \sqrt{N}\left( R^{(4)}+{\L\over 2M^3}+
{N^mN^n\over N}R^{(4)}_{mn}\right. \nonumber\\
& &\qquad - \left. {1\over
4N}\left[g'_{mn}(g^{mn})'+(g^{mn}g'_{mn})^2\right]\right) 
+{\mathcal{L}}_{high} \label{Lgen}\\
{\mathcal{L}}_{high}\ &=&\ {\sqrt{N}\over
2N^2}\left(g^{mn}N^sN_{s;n}(\f^2)_{,m}- N^s_{\ ;s}N^m(\f^2)_{,m}
\right.  \nonumber\\
& &\qquad + \left. 2N^mN^n\left[N^k_{\ ;k}N_{m;n}-N^k_{\
;m}N_{k;n}\right]\right) \label{Lhigh}\\ 
{\mathcal{L}}_i\ &=&\ {{\mathcal{V}}_i\over 2M^3} \delta(t-t_i)
\label{Lbound}
\end{eqnarray}
In the above expression, $R^{(4)}_{mn}$ is the Ricci tensor
constructed from the metric $g_{mn}$, and a semicolon denotes
the covariant derivative with respect to this metric. Because
of the dependence on the fifth coordinate, all these operations
are to be carried out for fixed values of $t$. The tensor nature of
each of the terms, 
with respect to $4D$ coordinate transformations, can
be read off easily.

Here, we would like to emphasize the fact that, although our starting 
theory was manifestly invariant under $5D$ coordinate transformations,
only a residual $4D$ symmetry survives in (\ref{actiongen}) because of
the special choice of coordinates made. Thus,  action
(\ref{actiongen}) is invariant under
arbitrary $4D$-coordinate reparameterizations,
$x\to x'(x)$. Let us also note
that, since all components
of the metric, $G_{MN}$, are even functions of the
extra coordinate, it follows  from the chain rule, $F'(\t)=\dot{F}(\t)
\t'$, that expressions linear in ${d\over dt}$
derivative are odd and, therefore, do not contribute to the action.
We have omitted such expressions in (\ref{Lgen}). Furthermore, we have
also dropped surface terms arising from integration by parts. This,
therefore, gives the complete action involving the graviton (and its
KK tower) and the radion.

\section{Background Method}

To study the solutions of this theory systematically, we use the
background field  method 
\cite{H,TV} (see in particular section 4 in \cite{TV}). We write
each of  the field
variables as a sum of a classical (possibly large) background and a small
fluctuation, and expand the action up to quadratic terms in the
fluctuations.  For the metric we write
\begin{eqnarray}
g_{mn}(x,\t)\ &=&\ \tilde{g}_{mn}(x,\t)\ +\ h_{mn}(x,\t)\nonumber\\
g^{mn}(x,\t)\ &=&\ \tilde{g}^{mn}(x,\t)\ -\ h^{mn}(x,\t)\ +\
h^m_{\ k}(x,\t)h^{kn}(x,\t) \label{gback}
\end{eqnarray}
while, for the radion, we use the decomposition
\begin{equation}\label{radback}
\f(x)\ =\ r(x)\ +\ r(x)d(x)
\end{equation}
where $\tilde{g}_{mn}(x,\t)$ and $r(x)$ are the background fields
which are assumed to satisfy the classical field equations,
while $h_{mn}$ and $d$ are the corresponding fluctuations (note
that we have introduced $d$ as a dimensionless field). The new
(background) metric satisfies 
$\tilde{g}_{mk}\tilde{g}^{kn}=\delta_m^n$, and from now on all
indices are lowered and raised with the background metric
$\tilde{g}_{mn}$ and its inverse. For simplicity, we will refer to this
as  the metric.

Our next task is to decompose the action up to second order in $h$
and $d$. The zeroth order action leads to the classical
equations for the background fields. The linear terms vanish, since
the backgrounds satisfy the classical equations and, therefore, the
first nontrivial term in the expansion of the action corresponds to the
quadratic part of the action in the $h$ and the $d$ fields. We note
here that our interest lies in the case where the classical solutions
lead to $r(x) = r_{c} = {\rm constant}$ primarily for two
reasons. Such a case would correspond to have the highest symmetry and 
the calculations will be much simpler. 
When $r(x)$ is a constant, it is easy to see that the
terms in (\ref{Lhigh}) would contain terms that are at least third
order in $d(x)$ and, therefore, will not be relevant for our
analysis. It is for this reason that we have separated out these
terms in (\ref{Lgen}) and we will neglect this term in the rest of our
discussions. On the other hand, if non-constant solutions for $r(x)$ are
of interest, then, the term in (\ref{Lhigh}) will contribute to
the expansion up to quadratic order.

Before carrying out the expansion, however, let us
discuss the Israel junction conditions \cite{I} that are crucial in analyzing
solutions in this theory. Let us note that we are considering a theory
with singular sources and, in such a case, boundary conditions on the branes
are extremely important in determining solutions. The junction conditions for 
the metric $g_{mn}$, can be derived following \cite{JC}, and in the 
present case of an orbifold read as
\begin{equation}\label{bc}
\dot{g}_{mn}(x,\t)\vert_{t=t_i}\ =\ -\xi(t){{\mathcal{V}}_i\over 12
M^3} \sqrt{N(x,\t)} g_{mn}(x,\t)\vert_{t=t_i}
\end{equation}
where
\begin{eqnarray}
 \xi(t) = \left\lbrace
           \begin{array}{ccc}
               1 & {\rm when}  & t=0,    \\
              -1 & {\rm when}  & t=\pi.  \\
           \end{array}
         \right.
\end{eqnarray}
Decomposing the metric and the radion as in (\ref{gback}),
(\ref{radback}) and using (\ref{N}) as well as the results from appendix {\bf A}, 
we get order by order in the fluctuations
\begin{eqnarray}
\dot{\tilde{g}}_{mn}(x,\t)\vert_{t=t_i}\ &=&\
-\xi(t){{\mathcal{V}}_ir_c\over 12 M^3}
\tilde{g}_{mn}(x,\t)\vert_{t=t_i}\label{bcg}\\
\dot{h}_{mn}(x,\t)\vert_{t=t_i}\ &=&\
-\xi(t){{\mathcal{V}}_ir_c\over 12 M^3} \left( h_{mn}(x,\t)
+d(x)\tilde{g}_{mn}(x,\t)\right) \vert_{t=t_i}\label{bch}
\end{eqnarray}
and so on. The equations following from the action (\ref{actiongen})
need  to be solved subject to these boundary conditions.

\subsection{Classical Action}

The zeroth order action (which does not contain $d$ or $h$) is
obtained from (\ref{actiongen}) with $g$ and $\f$
replaced by $\tilde{g}$ and $r$ respectively. The equations of
motion for the two fields can be obtained by varying with respect
to $\tilde{g}$ and $r$ which leads to two equations - one tensor and
one scalar. Let us note parenthetically that $\delta G_{m5}$ (see
(\ref{metric})) is not independent and, consequently, the $m5$ Einstein
equation would not be independent.

The zeroth order action or the classical action is easily obtained
from (\ref{actiongen}) to have the form
\begin{eqnarray}
S^{(0)}\ &=&\ -2M^3\int d^4x\int_{-\pi}^\pi dt\sqrt{-\tilde{g}}
\left( r \widetilde{R}+r{\L\over 2M^3}\right.    \nonumber \\
& &\ -\left. {1\over 4r}\left[\tilde{g}'_{mn}(\tilde{g}^{mn})'
+(\tilde{g}^{mn}\tilde{g}'_{mn})^2\right]+
\sum_{i=0,\pi}{{\mathcal{V}}_i\over 2M^3}
\delta(t-t_i)\right)\label{zero}
\end{eqnarray}
The Euler-Lagrange equations, following from this action, have the
form (upon setting $r(x)=r_{c}$)
\begin{eqnarray}
0\ &=&\ \widetilde{R}+{\L\over 2M^3} + {1\over
4r^2_c}\left[\tilde{g}'_{mn}(\tilde{g}^{mn})'+(\tilde{g}^{mn}
\tilde{g}'_{mn})^2\right]\label{vacuumsprime}\\
0\ &=&\ \widetilde{R}_{mk}-{1\over 2}\widetilde{R}\ \tilde{g}_{mk}
-{\L\over 4M^3}\tilde{g}_{mk} + {1\over
8r^2_c}\Big[4\tilde{g}''_{mk}-4\tilde{g}^{ab}\tilde{g}''_{ab}
\tilde{g}_{mk} \nonumber\\
& &\ -  4\tilde{g}^{ab}\tilde{g}'_{am}\tilde{g}'_{bk}
-3(\tilde{g}^{ab})'\tilde{g}'_{ab}\tilde{g}_{mk} +
2\tilde{g}^{ab}\tilde{g}'_{ab}\tilde{g}'_{mk}-
\left(\tilde{g}^{ab}\tilde{g}'_{ab}\right)^2
\tilde{g}_{mk}\Big]\nonumber\\
& &\ - \sum_{i=0,\pi} {{\mathcal{V}}_i\over 4r_c
M^3}\delta(t-t_i)\tilde{g}_{mk} \label{vacuumtprime}
\end{eqnarray}

As expected, equation (\ref{vacuumtprime}) contains singular
terms proportional to $\delta$-functions. However, let us note that
these equations, which are written in terms of derivatives with
respect to $t \in [-\pi,\pi]$,  can be
simplified if we express them in terms of the variable $\t \in
[0,\pi]$ (Recall that all the field variables depend on $\tau$.). We note that
\[
\tilde{g}'_{mn}(\tau) = \dot{\tilde{g}}_{mn} \tau',\qquad
\tilde{g}''_{mn} (\tau) = \ddot{\tilde{g}}_{mn} + \dot{\tilde{g}}_{mn}
\tau''
\]
Since $\t'^2=1$, in the classical equations, we can simply replace
terms  quadratic in
single $t$ derivatives (primes) by those with $\tau$ derivatives
(dots). However, in terms with double $t$ derivatives, the change to
$\tau$ derivatives would introduce new delta function singularities
because of the $\tau''$ term. On the other hand, we note from
(\ref{bcg}) that $\dot{\tilde{g}}_{mn}$, on both the branes, is proportional
to  $\tilde{g}_{mn}$ so that these new singular terms cancel
precisely the singular delta function terms already present in the
equation (\ref{vacuumtprime}). In fact, 
this is a general result, namely, singular terms arising from changing
``primes'' to ``dots'' exactly cancel the singular boundary
terms that appear explicitly in the equations of motion, because of the
junction conditions. We would, however, like to stress that, although
the dot derivatives appear to be the natural ones in the equations
of motion, it is more useful to have the prime derivatives in the
action to avoid subtleties in integration by parts.
 
Rewritten in terms of the $\tau$ (dot) derivatives, the
equations of motion have the form
\begin{eqnarray}
0\ &=&\ \widetilde{R}+{\L\over 2M^3} + {1\over
4r^2_c}\left[\dot{\tilde{g}}_{mn}\dot{\tilde{g}}^{mn}+(\tilde{g}^{mn}
\dot{\tilde{g}}_{mn})^2\right]\label{vacuumsdot}\\
0\ &=&\ \widetilde{R}_{mk}-{1\over 2}\widetilde{R}\ \tilde{g}_{mk}
-{\L\over 4M^3}\tilde{g}_{mk} + {1\over
8r^2_c}\left[4\ddot{\tilde{g}}_{mk}-4\tilde{g}^{ab}\ddot{\tilde{g}}_{ab}
\tilde{g}_{mk}   \right. \nonumber\\
& &\ - \left. 4\tilde{g}^{ab}\dot{\tilde{g}}_{am}\dot{\tilde{g}}_{bk}
-3\dot{\tilde{g}}^{ab}\dot{\tilde{g}}_{ab}\tilde{g}_{mk} +
2\tilde{g}^{ab}\dot{\tilde{g}}_{ab}\dot{\tilde{g}}_{mk}-
\left(\tilde{g}^{ab}\dot{\tilde{g}}_{ab}\right)^2\tilde{g}_{mk}
\right]\label{vacuumtdot}
\end{eqnarray}
without any singular terms involving $\delta$-functions. These are
highly  nonlinear equations which
are  clearly nontrivial to solve in general. Therefore, we look for a solution
of the metric, $\tilde{g}_{mn}$, in the factorizable form,
\begin{equation}\label{factorize}
\tilde{g}_{mk}(x,\t)=a(\t)\ \underline{g}_{mk}(x)
\end{equation}
Requiring $\ \underline{g}_{mk}(x)\ \underline{g}^{kn}(x)\ =\
\delta_m^n\ $, we have
\begin{equation}
\tilde{g}^{mk}\ =\ {1\over a}\ \underline{g}^{mk},\qquad
\widetilde{R}_{mk}\ =\ \underline{R}_{mk},\qquad
\widetilde{R}={1\over a}\ \underline{R} \label{properties}
\end{equation}
Using (\ref{factorize}) and (\ref{properties}), equations
(\ref{vacuumsdot}) and (\ref{vacuumtdot}) respectively take the forms
\begin{eqnarray}
0\ &=&\ \underline{R}+a\left[ {\L\over 2M^3} +{3\over r^2_c}
\left({\dot{a}\over a}\right)^2 \right]\label{vacuumi}\\
0\ &=&\ \underline{R}_{mk}-{1\over 2}\ \underline{R}\ \
\underline{g}_{mk}+{\l\over 4}\ \underline{g}_{mk}
\label{vacuumii}
\end{eqnarray}
where we have defined
\begin{equation}\label{lambda}
\l \ =\ -a\left({\L\over M^3} + {6\over r^2_c}{\ddot{a}\over a}\right)
\end{equation}
Contracting (\ref{vacuumii}) with $\underline{g}^{mk}$, we obtain
\begin{equation}\label{curvature}
\underline{R}=\l
\end{equation}
The left-hand-side of (\ref{curvature}) is a function of $x$ only,
while the right hand side depends only on $\tau$. It follows, therefore,
that $\l={\rm constant} = \underline{R}$, which, in turn, implies that
the  metric $\underline{g}_{mk}$ defines an Einstein space
\begin{equation}\label{Einstein}
\underline{R}_{mk}\ =\ {\l\over 4}\ \underline{g}_{mk}
\end{equation}
with constant scalar curvature $\l$. It follows now 
from (\ref{indmetric}) and (\ref{properties}) that the induced
metric on each of the branes will define an Einstein space as well, with
curvature $\widetilde{R}(t=t_i)\ =\ {\l\over a(t=t_i)}$. We also note
that every solution, $\underline{g}_{mk}$, of (\ref{Einstein})
corresponding to a given $\l$, will have the same warp factor $a$
determined from (\ref{vacuumi}) and (\ref{lambda}).

The solutions for the warp factor $a$ are already known. The flat case,
$\underline{g}_{mk}=\eta_{mk}$, was analyzed in \cite{RSfirst},
while the dS and the AdS solutions were derived in
\cite{dSAdS}. It is clear  from the preceding discussion that these
solutions exhaust all possible cases for the warp factor and here we
will  briefly describe
an alternate and unified derivation of these solutions for completeness.

The equations (\ref{vacuumi}) and (\ref{lambda}) determining the warp
factor can be rewritten in  a more convenient form
\begin{eqnarray}
&&\left({\dot{a}\over a}\right)^2\ =\ -{r_c^2\over 3}\left(
{\L\over 2M^3} +{\l\over a}\right)\label{adot}\\
&&\ddot{a} a -(\dot{a})^2\ =\ {\l r_c^2\over 6}\ a \label{addot}
\end{eqnarray}
We note from the above equations that a constant solution for
$a(\tau)$ exists only for the physically uninteresting case when
$\Lambda = 0 = {\cal V}_{i}$, which we will not consider. When $a\neq
{\rm constant}$, equation (\ref{addot}) follows from (\ref{adot}),
and, therefore, this is the only equation that we need to analyze.
Furthermore, this equation must be supplemented by the boundary conditions
(see (\ref{bcg}))
\begin{equation}\label{bca}
\dot{a}\vert_{t=t_i}\ =\ -\xi(t){{\mathcal{V}}_ir_c\over 12 M^3}\
a\vert_{t=t_i}
\end{equation}

The general solution of (\ref{adot}) is of trigonometric type for
$\L>0$, of exponential type for $\L<0$ and of polynomial type for
$\L=0$. Keeping in mind the spirit of the RS model, where the aim is
to solve  the hierarchy problem without introducing large numbers, we
will consider only the case $\L<0$ here. Following \cite{RSfirst}, we
define 
\begin{equation}\label{kconst}
{\L\over 24M^3}\ =\ -k^2
\end{equation}
and parameterize the tensions on the branes, for convenience, as
\begin{equation}\label{alphabeta}
{{\mathcal{V}}_0\over 24M^3}\ =\ \alpha k,\qquad
{{\mathcal{V}}_\pi\over 24M^3}\ =\ -\beta k
\end{equation}
where $\alpha,\beta$ are arbitrary parameters. In terms of these
parameters, the equation for $a$ reads
\begin{equation}
\left({\dot{a}\over a}\right)^2\ =\ 4k^2r_c^2-{\l r_c^2\over
3a}\label{eqa}
\end{equation}
which needs to be solved subject to the boundary conditions
\begin{equation}\label{bcanew}
{\dot{a}\over a}(\t=0)\ =\ -2\alpha kr_c,\qquad {\dot{a}\over
a}(\t=\pi)\ =\ -2\beta kr_c
\end{equation}

The solutions of (\ref{eqa}) subject to (\ref{bcanew}) are easily
determined to be
\begin{eqnarray}
 a(\t) = \left\lbrace
           \begin{array}{ccc}
               {\l\over 12k^2} \cosh^2(kr_c(p-\t)) & {\rm when}  & \l>0,    \\
                          \exp(-2kr_c\t))  & {\rm when}  & \l=0,    \\
               {|\l|\over 12k^2} \sinh^2(kr_c(p-\t)) & {\rm when}  &
               \l<0,\label{solution}
           \end{array}
         \right.
\end{eqnarray}
When $\l\neq 0$, $r_c$ is determined completely in terms of $\alpha$
and $\beta$ as
\begin{equation}\label{rc}
kr_c\ =\ {1\over \pi}{\rm arcth} \left({\alpha-\beta\over
1-\alpha\beta}\right)
\end{equation}
and the parameter $p$ is defined through
\begin{eqnarray}\label{p}
              e^{2kr_c p}\ =\  {1+\alpha\over 1-\alpha}\quad  & {\rm
              when}&\quad  \l>0, \\
              e^{2kr_c p}\ =\  {\alpha+1\over \alpha-1}\quad  & {\rm
              when}&\quad 
              \l<0, 
\end{eqnarray}
The range of the parameters, $\alpha$ and $\beta$, can be obtained from 
three consistency relations. We note that $r_c>0$, which
through (\ref{rc}) implies that $0< {\alpha-\beta\over
1-\alpha\beta}<1$. Furthermore, the right-hand-side of
(\ref{p}) must be positive, so that $|\alpha|<1$ for $\l>0$ and
$|\alpha|>1$ for $\l<0$. Finally, we have 
$a(\t)>0$ for any $\t$ and it follows from (\ref{solution}) that
$p\notin [0,\pi]$ when $\l<0$. It is easy to show that, for the
solutions in (\ref{solution}), the right-hand-side of (\ref{eqa}) is
non-negative for any $\t\in[0,\pi]$ and, consequently, this does not
introduce any further restriction on the parameters $\alpha$ and $\beta$. The
limiting case $\l=0$ cannot be obtained from these and a solution
exists only if the two tensions satisfy $\alpha=\beta=1$, as can be
seen from (\ref{solution}) and (\ref{bcanew}). In this case, the value
of $r_c$ remains undetermined.

\subsection{Quadratic Action}

If the background fields satisfy the classical equations, then, the
linear order terms in the expansion of the action vanish, which can be
explicitly checked. The leading nontrivial correction to the action
comes at the quadratic order. After some algebra, the part of the
action (\ref{actiongen}), quadratic in $h_{mn}$ and $d$, can be
determined to be (we give some useful relations for this expansion
in appendix {\bf A})
\begin{equation}\label{quadratic}
S^{(2)}\ =\ -2M^3 r_c \int d^4x \sqrt{-\underline{g}}
\int_{-\pi}^\pi dt\left( L(d)+L(d\underline{h}) + L(\underline{h}) \right)
\end{equation}
where
\begin{eqnarray}
L(d)\ &=&\ {\l\over 4} r^2_c\t^2 d^{,m}d_{,m} -{3\over
r_c^2}(a')^2 d^2 \label{Ld}     \\ 
L(d\underline{h})\ &=&\ a\left( \underline{h}^{mk}_{~~;k} -
\underline{h}^{,m}\right) d_{,m} - 
{a\l\over 4} d\underline{h} +{3\over 2}{aa'\over r^2_c}
d\underline{h}'  \label{Ldh}   \\
L(\underline{h})\ &=&\ {a\over 2}\left( {1\over
2}\underline{h}_{ak;m}\underline{h}^{ak;m}-\underline{h}_{ak;m}
\underline{h}^{am;k}+\underline{h}^{mk}_{~~;k}\underline{h}_{,m} 
-{1\over 2}\underline{h}^{,m}\underline{h}_{,m}\right) \label{K} \\
& &\ + {a^2\over 4r_c^2} \left( \underline{h}'_{mk}(\underline{h}^{mk})'
- (\underline{h}')^2\right) 
+{3\over 4}{aa'\over r^2_c}\left
(
2\underline{h}'_{mk}\underline{h}^{mk}-\underline{h}'\underline{h}\right)
\label{Mi}  \\ 
& &\ + \left( {a\l\over 8}+{3\over 4}{(a')^2\over r_c^2}\right)
\left(2\underline{h}_{mk}\underline{h}^{mk}-\underline{h}^2 \right)
\label{Mii}  \\ 
& &\ - \sum_{i=0,\pi} {a^2\over r_c}{ {\mathcal{V}}_i\over 16M^3}
\left(2\underline{h}_{mk}\underline{h}^{mk}-\underline{h}^2\right)
\delta(t-t_i) \label{LB} 
\end{eqnarray}
Keeping the factorization of the background metric in mind, we have
introduced a new field $\underline{h}_{mk}(x,\t)$ through
\begin{equation}\label{hbardef}
h_{mk}=a \underline{h}_{mk}~~~;~~~ h^{mk}=a\underline{h}^{mk}~~~;~~~
\underline{h}=\underline{h}_{mk}\underline{g}^{mk}
\end{equation}
All indices  are raised and lowered with the background metric
$\underline{g}_{mk}$, which also defines the  covariant
derivatives. As a result, let us note that $4D$ covariant
derivatives simply commute with derivatives with respect
to $t$ (or $\t$). 

The boundary condition satisfied by
$\underline{h}_{mk}$ is determined from (\ref{bch}) to be
\begin{equation}\label{bchbar}
\dot{\underline{h}}_{mk}(x,\t)\vert_{t=t_i}\ =\ -\xi(t){{\mathcal{V}}_i
r_c\over 12 M^3} \ d~\underline{g}_{mk}\vert_{t=t_i}
\end{equation}
The most important feature to note from (\ref{quadratic}) is the 
mixing between the fields $\underline{h}_{mk}$ and $d$. In order to
understand the physical spectrum of the theory, we must, of course,
diagonalize the second order Lagrangian density. Unfortunately, this is
not the complete story since there is mixing between the two fields
arising from the boundary conditions (\ref{bchbar}) as well. 
Therefore, we need to diagonalize simultaneously the quadratic
Lagrangian density,
(\ref{quadratic}), as well as the junction conditions to linear order,
(\ref{bchbar}). We note that higher order mixing, both in the action
as well as in the junction conditions, would simply correspond to
higher order interactions and is not of interest for our analysis.

The presence of mixing, in the quadratic action, simply corresponds to
the fact that $\underline{h}_{mk}$ and $d$ are not the appropriate
field variables and the physical fields will be, in general, a
linear combination of the two. However, we note that, unlike
$\underline{h}_{mk}$, the field $d$, in some sense, carries a
direct physical meaning - it labels the deviation of the distance
between the two branes from the classical value $r_c$. Consequently, 
we prefer to treat the field $d$ as physical and look for a general
redefinition of the metric fluctuation. The most general linear
redefinition of $\underline{h}_{mn}$ can be seen to be
\begin{equation}\label{bcdiagon}
\underline{h}_{mk}(x,\t)\ =\ X_{mk}(x,\t) +(\ln a + f(\t))d(x)\ \u{g}_{mk}
+S(\t)d_{;mk}+Z(\t)d^{;m}_{\ ;m}(x)\ \u{g}_{mk}
\end{equation}
where compatibility with (\ref{bchbar}) requires (see also (\ref{bca}))
\begin{eqnarray}
\dot{X}_{mk}(x,\t)\vert_{t=t_i}\ & = &\ 0 \label{bcX}\\
 \dot{f}(\t)
\vert_{t=t_i}\  & = &\ \dot{S}(\t)\vert_{t=t_i}\ =\
\dot{Z}(\t)\vert_{t=t_i}\ \label{bcfSK}
\end{eqnarray}
Note that the term $L(\underline{h})$ is form-invariant under the
transformation (\ref{bcdiagon}). However, $L(d)$ and
$L(d\underline{h})$ do transform  and it is easy
to see  that if $Z\neq0$, such a transformation will generate mixing
between $X_{mk}$ and $d$ with fourth order derivatives. Since this
does not help in diagonalization, we set $Z=0$.  Furthermore, we find
that there is no such $f$ and $S$  obeying
(\ref{bcfSK}) which set the mixing terms, in the Lagrangian density,
to zero.

There is another possible solution to the mixing problem. We note here
that it is the
action that we need to diagonalize (subject to the boundary
conditions). Therefore, if we
know the KK decomposition for the field $X_{mk}$, then the action
(\ref{quadratic}) would be a $5D$ action for $4D$ fields that
depend only on the coordinate $x$, with $t$-dependent
coefficients. It would, therefore, be  sufficient to
arrange the coefficient functions in front of the mixing terms
to vanish when integrated over the extra dimension so that such terms
can be removed from the action.

To find a set of basis functions, in terms of which we can decompose the
field $X_{mk}$, we proceed as follows. Let us assume that the
change (\ref{bcdiagon}) leads to a diagonalization of the action (it
also diagonalizes the boundary conditions because of
(\ref{bcX})). Since the transformation (\ref{bcdiagon}) is linear and
leaves  the part of the action depending only on $\underline{h}_{mk}$
form-invariant, the other two terms in the action can only lead to terms
that are quadratic in the field $d$. As a result, the KK decomposition of
$X_{mk}$ can be easily studied by analyzing the $\underline{h}_{mk}$ part of
the action and  we can simply set $d=0$ in
(\ref{quadratic}) and (\ref{bchbar}) for this purpose. Having the
basis  functions determined in this way,
$X_{mk}$ can be expanded in this basis ($d(x)$ does not have a KK
tower)  in the action
(\ref{quadratic}), and the resulting action diagonalized 
at each level of the KK tower. As we will see, this 
procedure works quite nicely. We discuss the actual solutions of
(\ref{quadratic}) and (\ref{bchbar}) for $d=0$ in appendix {\bf B}.

The most convenient way to implement the program discussed above, is
to first make the change of variables
\begin{equation}\label{changeone}
\underline{h}_{mk}(x,\t)\ =\ X_{mk}(x,\t) +(\ln a + f(\t))d(x)\ \u{g}_{mk}
\end{equation}
with
\begin{equation}\label{bcXone}
\dot{X}_{mk}(x,\t)\vert_{t=t_i}\ =\ \dot{f}(\t)\vert_{t=t_i}\ =\ 0
\end{equation}
The function $f(\t)$ is otherwise arbitrary. Since the new field $X_{mk}$
satisfies (\ref{bcXone}), we can decompose it in the system of
basis functions $\F^\a$ (whose derivative vanishes at the boundaries,
see  appendix {\bf B}  for details) as
\begin{equation}\label{decompPhi}
X_{mk}(x,\t)\ =\ \sum_\a\  \AX_{mk}(x) \F^\a(\t)
\end{equation}
The action (\ref{quadratic}) can now be written in the 
form
\begin{equation}\label{quadraticX}
S^{(2)}\ =\ -2M^3 r_c \int d^4x \sqrt{-\underline{g}}
\int_{-\pi}^\pi dt\left( L(d)+L(dX) + L(X) \right)
\end{equation}
where
\begin{eqnarray}
L(d)\ &=&\ K(\t)\ d^{,m}d_{,m}\ +\ {M(\t)\over r_c^2}\ d^2 \label{LX(d)}     \\
L(dX)\ &=&\ a(1+\ln a + f)\left[ \left( X^{mk}_{~~;k} -
X^{,m}\right) d_{,m} - {\l\over 4} dX\right]\ +\ {3a^2\over
2r_c^2}\left(\ddot{f}+2{\dot{a}\over a}\dot{f}\right) dX \label{L(dX)}   \\
L(X)\ &=&\ {1\over 2}\sum_\a\ a\left(\F_\a \right)^2\
L_{PF}(\AX_{mk};m^2_\a) \label{L(X)}
\end{eqnarray}
Here we have defined $K$ and $M$ as
\begin{eqnarray}
K(\t)\ &=&\ {\l\over 4}r_c^2\t^2-{3\over 2}a(\ln a+f)(\ln a+f+2)\label{Kin}     \\
M(\t)\ &=&\ -6a\ddot{a}(\ln a+f)(\ln a+f+1)\ -\ 3a\dot{f}(
a\dot{f}+2\dot{a}(\ln a+f)) \nonumber\\
& &\ + \sum_{i=0,\pi}~a^2{{\mathcal{V}}_ir_c\over 2M^3} (\ln
a+f)^2\delta(t-t_i)\label{Mass}
\end{eqnarray}
and $L_{PF}$ represents the Pauli-Fierz term that we will discuss
later (it is also discussed in appendix {\bf B}).

Because of (\ref{bcXone}), we can also decompose $f$ in terms of
$\F^\a$ as
\begin{equation}\label{decompf}
f(\t)\ =\ \sum_\a\  f_\a \F^\a(\t)
\end{equation}
Using (\ref{eqPhi}) and (\ref{scalprod}), we can rewrite the mixing terms,
(\ref{L(dX)}), in the action in the form 
\begin{eqnarray}
L(dX)\ &=&\ a(1+\ln a + \rho)\F^0\left[ \left( \OX^{mk}_{~~;k} -
\OX^{,m}\right) d_{,m} - {\l\over 4} d\OX\right]\nonumber\\
& &\ + \sum_{\a\geq 1} a(\F^\a)^2\left[ (I^\a+f_\a)\left( \left(
\AX^{mk}_{~~;k} - \AX^{,m}\right) d_{,m} - {\l\over 4}
d\AX\right)\right.\nonumber\\ 
& &\qquad - \left.{3\over 2}f_\a m^2_\a d\AX \right] \label{L(dX)ii}
\end{eqnarray}
where we have introduced the notations
\begin{equation}\label{rho,I}
\rho\ =\ f_0\F^0\ =\ {\rm constant},\qquad I^\a\ =\ \int_{-\pi}^\pi
dt~ a\ln a\ \F^\a~,~ \a\geq 1
\end{equation}
and have used the equality $\int_{-\pi}^\pi
dt~ a\ln a\F^\a =\ \int_{-\pi}^\pi dt~ a (\F^\a)^2 I^\a .$

It is impossible to find a set of constants, $f_\a$, which will make
the mixing terms in the action vanish. As a result,
we are forced to make a second transformation involving the
second derivative of $d$, which is done at each level
of the KK tower
\begin{equation}\label{changetwo}
\AX_{mk}(x)\ =\ \Ah_{mk}(x)\ +\ \s_\a\ d_{;mk}~~~,~~~ \a\geq 0
\end{equation}
The numbers $\s_\a$ can be thought of as the coefficients in the
decomposition of the function $S(\t)$, in (\ref{bcdiagon}), in the basis
functions $\Phi^{\alpha}$, namely,
\begin{equation}
S(\tau) = \sum_{\alpha} \sigma_{\alpha} \Phi^{\alpha} (\tau)\label{S}
\end{equation}
Under
the transformation (\ref{changetwo}), the Pauli-Fierz Lagrangian
(see (\ref{PFlag})) becomes 
\begin{eqnarray}
L_{PF}(\AX,m^2_\a)\ &=&\ L_{PF}(\Ah,m^2_\a)\ -m^2_\a\s_\a\left(
\Ah^{mk}_{~~;k} - \Ah^{,m}\right) d_{,m}   \nonumber\\
& &\ - {\l\s^2_\a\over 8}\left( d^{;m}_{~~;m}\right)^2 + {\l\s^2_\a
m^2_\a\over 8}\ d^{,m}d_{,m}\label{changeLPF}
\end{eqnarray}
and the mixing term (\ref{L(dX)ii}) takes the form
\begin{eqnarray}
L(dX)\ &=&\ a(1+\ln a + \rho)\F^0\left[ \left( \Oh^{mk}_{~~;k} -
\Oh^{,m}\right) d_{,m} - {\l\over 4} d\Oh\right]\nonumber\\
& &\ + \sum_{\a\geq 1} a(\F^\a)^2\left[ (I^\a+f_\a
-{m^2_\a\s_\a\over 2} )\left( \Ah^{mk}_{~~;k} - \Ah^{,m}\right)
d_{,m}\right.\nonumber\\
& &\qquad + \left. \left( -{\l\over 4}(I^\a+f_\a)-{3\over 2}f_\a
m^2_\a\right) d\Ah\right]\label{L(dH)ii}
\end{eqnarray}
The mixing for $\a=0$ can be removed if we choose $\rho$ to satisfy
\begin{equation}\label{rho}
\int_{-\pi}^\pi dt~a(1+\ln a + \rho)\ =\ 0
\end{equation}
For $\a\geq 1$, there are two conditions that must hold for the mixing to
disappear, namely,
\begin{eqnarray}
 I^\a + f_\a\ & = &\ {m^2_\a \s_\a\over 2}\nonumber\\
 {\l\over 4}(I^\a+f_\a)\ +\ {3\over 2}f_\a m^2_\a\ & = &\
0\nonumber\
\end{eqnarray}
which can be easily solved to give
\begin{eqnarray}
f_\a\ &=&\ -{\l\over \l+6m^2_\a}I^\a\label{f}\\
\s_\a\ &=&\ {12 \over \l+6m^2_\a}I^\a \label{sigma}\
\end{eqnarray}
(Note that the above expressions are free from singularities in
view of (\ref{inequality}).)

Thus, we see that the transformation (\ref{changeone}) together with 
(\ref{changetwo}) diagonalizes the quadratic action as well as
the boundary conditions, with $f(\t)$ determined from  (\ref{rho}) and
(\ref{f}). We note that, while $\s_{\a}$, for $\a\geq 1$, is determined from
(\ref{sigma}) the
coefficient $\s_0$ remains arbitrary. (This
arbitrariness is related to the invariance of the theory under a
gauge transformation of the graviton of the form (\ref{gaugetransf}).)

The diagonalized action now takes the form
\begin{equation}\label{quadraticH}
S^{(2)}\ =\ -2M^3 r_c \int d^4x \sqrt{-\underline{g}}
\int_{-\pi}^\pi dt\left( L(d) + L(H) \right)
\end{equation}
where
\begin{eqnarray}
L(d)\ &=&\ -{\l\over 16}\left( \sum_{\a\geq
0}~a(\F^\a)^2\s^2_\a\right)(d^{;m}_{~~;m})^2 \nonumber\\
& &\ + \left(K(\t)- {\l\over 16}\sum_{\a\geq 1} ~a(\F^\a)^2 m^2_\a
\s^2_\a\right) d^{,m}d_{,m}\ +\ {M(\t)\over
r_c^2}~d^2\label{LH(d)}\\
L(H)\ &=&\ {1\over 2}\sum_{\a\geq 0} a\left(\F^\a \right)^2\
L_{PF}(\Ah_{mk};m^2_\a) \label{L(H)}
\end{eqnarray}
and the functions $K$ and $M$ are defined in (\ref{Kin}) and
(\ref{Mass}). The properties of the diagonalized theory will be
discussed in more detail in the next section.

Let us present here the final structure of the metric $g_{mk}(x,\t)$
after all the transformations. Combining
(\ref{gback}), (\ref{factorize}), (\ref{hbardef}),
(\ref{changeone}) and (\ref{changetwo}) we have
\begin{eqnarray}
g_{mk}(x,\t)\ &=&\ a(\t)\left[ \left( 1+d(x)[\ln a~ +f(\t)]
\right)~\u{g}_{mk}(x) + \right. \nonumber\\
& &\ +\left.  \Oh_{mk}(x)\F^0 + S(\t) d_{;mk}(x) + \sum_{\a\geq 1}
\Ah_{mk}(x)\F^\a (\t)  \right] \label{finalmetric}
\end{eqnarray}
The functions $f(\t)$ and $S(\t)$ are given in terms of their
coefficients (\ref{rho}), (\ref{f}) and (\ref{sigma})
respectively. As we noted earlier, the function $S(\t)$ is
determined up to an additive constant. The reason is quite
clear. As we will discuss in appendix {\bf B}, the massless field
$\Oh_{mk}$ is  determined only up to a
gauge transformation (\ref{gaugetransf}) and, therefore, the term
proportional to $d_{;mk}$ in (\ref{bcdiagon}) can be gauged away at
any fixed  point
$t={\rm constant}$, in particular, on any one of the two branes, but
cannot  be gauged away in 
the whole bulk.

\section{4D Effective Theory}

To study further the properties as well as the physical implications of
the  diagonalized theory,
we need to know the values of the coefficients present in
(\ref{LH(d)}) and (\ref{L(H)}) after integration over the fifth
coordinate. Although so far we have treated both the
cases, $\l=0$ and $\l\neq 0$, on the same footing, it will be more
convenient, in this section, to consider these two cases separately.
\vskip 5mm\noindent {\bf\underline{${\bf \l= 0}$:}}\vskip 5mm
Let us note, at the outset, that $\l =0$ does not automatically imply
$\u{g}_{mk}=\eta_{mk}$. For an arbitrary Einstein space, we have
\cite{Weinberg}
\begin{equation}\label{Weil}
\u{R}_{makb}\ =\ -{\l\over 12}\left(
\u{g}_{mb}\u{g}_{ak}-\u{g}_{mk}\u{g}_{ab}\right) + \u{C}_{makb}
\end{equation}
where $\u{C}_{makb}$ is the Weyl conformal tensor, and 
the solutions for any given $\l$ are completely
degenerate with respect to different choices of $\u{C}_{makb}$.

When $\lambda = 0$, it is clear from (\ref{f}) that
$f_{\alpha}=0$ for $\alpha\geq 1$, so that $f(\t)=\rho$ and the warp factor
is given in (\ref{solution}). With these, a direct evaluation of (\ref{rho})
determines
\begin{equation}\label{valuerho}
\rho\ =\ -{2\pi k r_c\over e^{2\pi k r_c}-1}\ <\ 0
\end{equation}
The coefficient in front of the kinetic term of the radion can now be
determined from (\ref{LH(d)}) and (\ref{Kin}) to be
\begin{equation}\label{valueKin}
\kappa_d\ = \int_{-\pi}^{\pi} dt K(\tau)\ =\ -{3\over
2}\int_{-\pi}^\pi a(\ln a +\rho)(\ln a +\rho +2) dt\ =\ -3\pi\rho\ >\ 0
\end{equation}
Similarly, we can obtain the radion mass from (\ref{LH(d)}) and
(\ref{Mass}) and the explicit evaluation gives
$$
\int_{-\pi}^\pi a^2(\ln a+\rho)(\ln a+\rho+1) dt\ =\
\int_{-\pi}^\pi \sum_{i=0,\pi}~a^2{{\mathcal{V}}_ir_c\over 2M^3}
(\ln a+\rho)^2\delta(t-t_i) dt\ =\ 0
$$
Therefore, the radion is massless, since
\begin{equation}\label{valueMass}
\int_{-\pi}^\pi {M(\t)\over r_c^2} dt \ =\ 0
\end{equation}

The quadratic action, therefore, takes the form
\begin{equation}\label{effzero}
S_{eff}\ =\ \int d^4x \sqrt{-\u{g}}\ \left( -6\pi |\rho | M^3 r_c
d^{,m}d_{,m} -M^3 r_c\ \sum_{\a\geq 0}\
L_{PF}(\Ah_{mk};m^2_\a)\right)
\end{equation}
Let us next normalize the  fields $d$ and $\Ah_{mk}$ in the following
way. Let us assume that $\exp(-2\pi k r_c)\ll 1$  and express $M$ in
terms of  the $4D$ effective Planck mass
$M_{PL}^2=M^3/k $ \cite{RSfirst}. Furthermore, let us define 
\begin{eqnarray}
d(x)\ &=&\ {e^{k r_c\pi}\over \sqrt{24}\pi k r_c M_{PL}} D(x) \nonumber\\
\Ah_{mk}(x)\ &=&\ {1\over \sqrt{k r_c} M_{PL} }\APX_{mk}(x)
\label{normfields}
\end{eqnarray}
In terms of these fields, the quadratic action
becomes:
\begin{equation}\label{effzeronorm}
S_{eff}\ =\ \int d^4x \sqrt{-\u{g}}\ \left( -{1\over 2}
D^{,m}D_{,m} - \sum_{\a\geq 0}\ L_{PF}(\APX_{mk};m^2_\a)\right)
\end{equation}
The content of the theory, at the quadratic level, is now obvious - there is
a massless scalar field (the radion), a massless spin-2 field (graviton)
and an infinite tower of massive spin-2 fields with masses $m_\a^2$.
We see that no ghost fields are present in this theory, since the
kinetic term of the radion has the right sign, and the fields $\APX_{mk}$
carry precisely spin-2 content. We can, therefore, think of these
fields as the physical fields. The non-presence of (ghost) scalar
component in the fields $\APX_{mk}$ (for $\a\geq 1$) is due to the
conditions (\ref{gaugeH}) which these fields satisfy. Note that in
our framework, the conditions (\ref{gaugeH}) arise naturally from
the equations of motion for these fields (for more details see the
appendix in \cite{Veltman}), and do not have to be imposed by
hand.

To understand the coupling of the fields $D(x)$ and $\APX_{mk}$ to
matter, we need to express the metric $g_{mk}(x,\t)$ in terms of these
physical fields. To that end, let us recall the expression for the normalized
functions $\F^\a(\t)$ \cite{Rizzo}
\begin{equation}\label{Fnorm}
\F^\a(\t)\ =\ {\sqrt{k r_c}e^{k r_c \pi}\over J_2(x_\a)}
e^{2kr_c(\t-\pi)} \left[ J_2(x_\a e^{kr_c(\t-\pi)})+ {\pi\over 4}
x_\a^2 e^{-2kr_c\pi} Y_2(x_\a e^{kr_c(\t-\pi)})\right]
\end{equation}
for $\a\neq 0$, and $\F^0\ =\ \sqrt{kr_c}$. Here, $x_\a$'s represent
the positive zeroes of the Bessel function $J_1(x)$. Furthermore, the masses,
$m_\a^2$, of the fields $\APX_{mk}$ are given by \cite{Rizzo}
\begin{equation}\label{masses}
m_\a^2\ =\ k^2x_\a^2 e^{-2kr_c\pi}
\end{equation}
On the visible brane, we have
\begin{equation}\label{Fvis}
\F^\a(\t=\pi)\ =\ \sqrt{kr_c} e^{kr_c\pi}~~,~~ \a\geq 1
\end{equation}
while on the hidden brane they are rescaled roughly by a factor
($\a$-dependent) of the order of $\exp (-2kr_c\pi)$.  Collecting earlier
results, we can now write the metric $g_{mk}(x,\t)$ in
terms of the physical fields (\ref{normfields}) as
\begin{eqnarray}
g_{mk}(x,\t)\ &=&\ e^{-2kr_c\t}\left[ \left( 1-{(\pi
e^{-2kr_c\pi}+\t)e^{kr_c\pi}\over \sqrt{6}\pi M_{PL}} D(x)\right)
\u{g}_{mk}(x) + {S(\t) e^{kr_c\pi}\over \sqrt{24}\pi kr_c M_{PL} }
D_{;mk}(x) \right.\nonumber\\
& &\ +\left. {1\over M_{PL}}\OPX_{mk}(x) + {1\over \sqrt{kr_c}
M_{PL}} \sum_{\a\geq 1} \APX_{mk}(x)\F^\a (\t) \right]
\label{physmetric}
\end{eqnarray}
Of particular interest to us is the restriction of (\ref{physmetric}) to
the visible brane, which would determine the coupling of the fields
$D$ and $\APX_{mk}$ to matter located there.
\begin{eqnarray}
g_{mk}^{vis}(x)\ &=&\ e^{-2kr_c\pi}\left[ \left(
1-{e^{kr_c\pi}\over \sqrt{6} M_{PL}} D(x)\right) \u{g}_{mk}(x)
\right. \label{vismetricA}\\
& &\quad +\left. {1\over \sqrt{6\pi^2 kr_c}}{e^{3kr_c\pi}\over  k^2
M_{PL} } \left( e^{kr_c\pi}\sum_{\a\geq 1}{I^\a\over x_\a^2} \right)
D_{;mk}(x) \right.\label{vismetricB}\\
& &\quad +\left. {1\over M_{PL}}\OPX_{mk}(x) + {e^{kr_c\pi}\over M_{PL}}
\sum_{\a\geq 1} \APX_{mk}(x) \right] \label{vismetricC}
\end{eqnarray}
Some comments are in order. The couplings of the spin-2 states 
(\ref{vismetricC}) were already discussed in \cite{Rizzo}. The radion has 
two types of couplings to matter. The non-derivative one, on the visible 
brane, has strength comparable to that  of the massive spin-2 states in 
(\ref{vismetricC}) which is of the order of $(TeV)^{-1}$. 
This coupling is nonzero on the hidden brane as 
well, although much weaker (as also discussed in \cite{Rubakov}). The coupling on the 
hidden brane has its origin in the mixing of the radion with the graviton. 
The expression for the derivative coupling of the radion is nontrivial, due to the fact 
that we account for the presence 
of the massive KK tower of states. As we have noted earlier, the mixing of
the  radion, involving derivatives,  with the graviton in
(\ref{bcdiagon}) involves only the coefficient $\s_0$, which is a
constant and can, therefore, be gauged away in the whole bulk. However, the 
mixing of the radion involving derivatives with the massive KK tower in 
(\ref{bcdiagon}) involves coefficients that depend on $\t$, namely, the 
function $S(\t)$ in (\ref{physmetric}). As a result, in general, its value 
will be different at different points of the fifth dimension. The derivative 
coupling term can be arranged to vanish on the visible brane, but
unless the function $S(\t)$ takes special values, the term $D_{;mk}$ cannot 
be simultaneously gauged away on both the branes.  
Let us estimate the coefficient of this term in
(\ref{vismetricB}). Since $ k \lesssim M_{PL}$ (see \cite{Rizzo}),
the number outside the parenthesis is of the  order of $(TeV)^{-3}$.
The term inside the parenthesis needs a more careful
consideration. We use $kr_{c} = 12$ and take the values for the first fifty
roots, $x_{\alpha}$, with eight digit precision from \cite{Tables}. A
direct numerical evaluation shows that the first term of the series is
$-3.5593$.  The subsequent terms, in the series, are of alternating sign and
tend to have very slowly decreasing magnitude. We
believe that the series is absolutely convergent. The numerical
evaluation of the series including the first fifty terms yields the
value  $0.74$ times the first
term. This value essentially stabilizes after approximately the twentieth
term. The correction to this value from the remainder of the series is of
the order of the next term, which is of the order of
$10^{-3}$. This is roughly the order of magnitude of each of the terms
in the series for the next fifty or so terms. The $200$-th term, in the
series, for example, has a value of the order of $10^{-4}$. Therefore,
we estimate the term in the parenthesis to be of the order of unity. 

\vskip 5mm\noindent {\bf\underline{${\bf \l \neq 0}$:}}\vskip 5mm

When $\lambda \neq 0$, the diagonalized quadratic action is of the form
\begin{equation}
S_{eff}\ =\ \int d^4x \sqrt{-\u{g}}\ {\mathcal{L}}(d)+\sum_{\a\geq
0}\ \int d^4x \sqrt{-\u{g}}\ {\mathcal{L}}^\a \label{effnonzero}
\end{equation}
where
\begin{eqnarray}
{\mathcal{L}}(d)\ &=&\ \zeta_1\ (d^{;m}_{~~;m})^2 + \zeta_2\ 
d^{,m}d_{,m} + \zeta_m\  d^2\nonumber\\
{\mathcal{L}}^\a\ &=&\ -M^3 r_c L_{PF}(\Ah;m^2_\a)\nonumber\
\end{eqnarray}
and
\begin{eqnarray}
\zeta_1\ &=&\ {M^3r_c\l\over 8}\left( \sum_{\a\geq
0}\s_{\a}^2\right)\nonumber\\
\zeta_2\ &=&\ -2M^3r_c\left( \int_{-\pi}^\pi dt K(\t) -{\l\over
16}\sum_{\a\geq 1} m_\a^2 \s_\a^2\right)\nonumber\\
\zeta_m\ &=&\ -2{M^3\over r_c}\int_{-\pi}^\pi dt M(\t)\label{zeta}
\end{eqnarray}
The spin-2 content of the theory has properties similar to the case
$\l =0$  and
does not need any further discussion. However, the radion now 
exhibits new features. The kinetic term has a fourth
order derivative term whose coefficient has the sign of $\l$. Although in
principle the $\l\neq 0$ case can be analyzed in a fashion similar
to the $\l=0$ case, there are some technical complications which make
the analysis quite difficult. For example, we do not know of a
complete   system of
functions $\F^\a$ that is of a simple form, although equation
(\ref{eqPhi}) can be solved in terms of associated Legendre
functions or equivalently in terms of hypergeometric
functions. Since the
functions $K(\t)$ and $M(\t)$ (see (\ref{Kin}), (\ref{Mass})) are
given in  terms of the function
$f(\t)$, which in turn can only be determined from its series
expansion, we do not have much knowledge about these functions
either. Therefore, we are unable to discuss the values of the
coefficients (\ref{zeta}) any further. Nonetheless, it is easy to see that the
coefficient $\zeta_1$ is unlikely to be zero and has the same sign as
$\l$ (It will be zero if all the $I^{\alpha}$'s are to vanish which is
highly unlikely although not impossible.).
This is sufficient to indicate that higher derivative
ghosts will be associated with the radion field in this case.

\section{Conclusions}

In this paper, we have systematically studied, within the Lagrangian
approach,  the question of
identification and consistent inclusion of the radion field in the two
brane Randall-Sundrum model. We have exploited
the symmetries of the theory to identify the radion in an unambiguous
manner. Using the background field method, we have given a unified
derivation of the classical solutions as well as the form of the
action at quadratic order in the fields. We have shown that the
background metric, in general, corresponds to that of an Einstein
space with a warp factor. We have discussed in detail
the question of diagonalization of this quadratic action
which is essential in understanding the spectrum of the theory. We
have studied further the effective 4-dimensional action following from
this diagonalized action and it is
shown that the radion has no Kaluza-Klein tower and is massless for
backgrounds  with $\lambda = 0$. In this case, the graviton as well
as the  infinite
tower of states truly describe spin-2 particles and the theory is free
of ghosts. For the case $\lambda\not= 0$, however,
the  situation is
less clear and higher derivative ghost terms, involving the radion, appear
in  the action. The question of matter coupling  is
also discussed. We would like to emphasize here that there are only two
assumptions that we have made, i) the background
$r(x)=r_{c} = {\rm constant}$ and ii) a factorizable form of the
background metric, $\tilde{g}_{mn} (x,\tau) = a(\tau)
\underline{g}_{mn}(x)$. The entire analysis, otherwise, is quite general.

\vskip .7cm
\noindent{\bf Acknowledgement:}

This work was supported in part by USDOE Grant number DE-FG-02-91ER40685.

\appendix

\section{Some useful relations}

In this appendix, we collect some relations that are useful in the 
decomposition of the action as well as the junction conditions. For
$N$ defined  in (\ref{N}), and with
(\ref{gback}), (\ref{radback}), we have the following
decompositions up to second order in the fluctuations $h_{mn}$ and
$d$.
\begin{eqnarray}
N\ &=&\ r_c^2\left( 1+2d +d^2 -\t^2
r_c^2\ \tilde{g}^{ab}d_{,a}d_{,b} \right)\nonumber\\
\sqrt{N}\ &=&\ r_c\left( 1+d +d^2 -{\t^2 r_c^2\over 2}
\tilde{g}^{ab}d_{,a}d_{,b} \right)\nonumber\\
{1\over N}\ &=&\ {1\over r_c^2}\left(1-2d +3d^2 +\t^2
r_c^2\ \tilde{g}^{ab}d_{,a}d_{,b} \right)\nonumber\\
{1\over \sqrt{N}}\ &=&\ {1\over r_c}\left( 1-d +d^2 +{\t^2
r_c^2\over 2}\tilde{g}^{ab}d_{,a}d_{,b} \right) \label{Ndecomp}
\end{eqnarray}
The determinant of the metric $g_{mn}$ can also be 
decomposed up to quadratic order as (see \cite{TV})
\begin{equation}
\sqrt{-g}\ =\ \sqrt{-\tilde{g}}\left( 1+{1\over 2}h-{1\over 4}
h^a_{\ b}h^b_{\ a} + {1\over 8} h^2 \right) \label{gdecomp}
\end{equation}
where $h=\tilde{g}^{mk} h_{mk}$, and we have, to quadratic order in the
fluctuations, 
\begin{equation}
\sqrt{N}\sqrt{-g}\ =\ r_c\sqrt{-\tilde{g}}\left(  1+d +{1\over 2}h
+{1\over 2}dh + {1\over 8} h^2 -{1\over 4} h^a_{\ b}h^b_{\ a}
-{\t^2 r_c^2\over 2} \tilde{g}^{ab}d_{,a}d_{,b} \right)
\label{NGdecomp}
\end{equation}
The decomposition of $R^{(4)}_{mn}$ and $R^{(4)}$ can be found in
\cite{TV} (note that the notations there differ slightly from
ours).

\section{Kaluza-Klein Decomposition}

In this appendix, we describe the Kaluza-Klein decomposition for the
fluctuation of the metric, $H_{mk}$, and discuss some of the
properties of the basis functions. Let us consider the action
(\ref{quadratic}). When $d=0$, it is easy to check from
eqs. (\ref{changeone}) and (\ref{changetwo}) that
\[
h_{mk} = X_{mk} = H_{mk}
\]
so that the quadratic action takes the form
\begin{equation}\label{actionH}
S(H)\ =\ -2M^3 r_c \int d^4x \sqrt{-\underline{g}}
\int_{-\pi}^\pi dt\ L(H)
\end{equation}
where
\begin{eqnarray}
L(H)\ &=&\ {a\over 2}\left( {1\over
2}H_{ak;m}H^{ak;m}-H_{ak;m}H^{am;k}+H^{mk}_{~~;k}H_{,m}
-{1\over 2}H^{,m}H_{,m}\right) \label{HK} \\
&+& {a^2\over 4r_c^2} \left( H'_{mk}(H^{mk})' - (H')^2\right)
+{3\over 4}{aa'\over r^2_c}\left( 2H'_{mk}H^{mk}-H'H\right) \label{HMi}  \\
&+& \left( {a\l\over 8}+{3\over 4}{(a')^2\over r_c^2}\right)
\left(2H_{mk}H^{mk}-H^2 \right) \label{HMii}  \\
&-& \sum_{i=0,\pi} {a^2\over r_c}{ {\mathcal{V}}_i\over 16M^3}
\left(2H_{mk}H^{mk}-H^2\right) \delta(t-t_i) \label{HLB}
\end{eqnarray}
Here, as usual, $H = H^{mk}\underline{g}_{mk}$ and the field
$H_{mk}$ has to satisfy  the boundary condition (see (\ref{bchbar}))
\begin{equation}\label{bcH}
\dot{H}_{mk}(x,\t)\vert_{t=t_i}\ =\ 0
\end{equation}

Varying (\ref{actionH}) with respect to $H_{mk}$, we obtain 
\begin{eqnarray}
&& H_{mk;a}^{~~~~~;a}-H^{;a}_{\ ;a}\ \u{g}_{mk} +
H_{;mk}+H^{ab}_{~~;ab}\ \u{g}_{mk} -
H_{ma;k}^{~~~~~;a}-H_{ka;m}^{~~~~~;a} \nonumber \\
&&+{a\over r_c^2}\left(\ddot{H}_{mk}-\ddot{H}\ \u{g}_{mk}\right)
+2{\dot{a}\over r_c^2}\left(\dot{H}_{mk}-\dot{H}\
\u{g}_{mk}\right) -{\l\over 4}\left(2H_{mk}-H\
\u{g}_{mk}\right)\ =\ 0\label{eqHvd}
\end{eqnarray}
As we have seen from the discussion following (\ref{vacuumtprime}),
all singular terms from the equations of motion disappear when we
use dot ($\tau$) derivatives. The above equation admits separation of
variables and let us introduce a system of functions
$\F^\a(\t)$ such that:
\begin{equation}\label{Phi}
H_{mk}(x,\t)\ =\ \sum_\a\  \AH_{mk}(x) \F^\a(\t)
\end{equation}
Then, (\ref{eqHvd}) can be rewritten as
\begin{eqnarray}
&& \AH_{mk;a}^{~~~~~;a}-\AH^{;a}_{\ ;a}\ \u{g}_{mk} +\AH_{;mk}+
\AH^{ab}_{~~;ab}\ \u{g}_{mk} -
\AH_{ma;k}^{~~~~~;a}-\AH_{ka;m}^{~~~~~;a} \nonumber \\
&&\;\;-{\l\over 4}\left(2 \AH_{mk}- \AH\ \u{g}_{mk}\right)
-m^2_\a\left( \AH_{mk}- \AH\ \u{g}_{mk}\right) \ =\ 0
\label{eqHA}\\
&&   \nonumber\\
&& \ddot{\F}^\a + 2{\dot{a}\over a}\dot{\F}^\a +
{m^2_\a r_c^2\over a} \F^\a\ =\ 0\label{eqPhi}
\end{eqnarray}
where $m^2_\a$ represent the separation constants.  

Let us first analyze the system of functions $\F^\a(\t)$, defined
through  equation (\ref{eqPhi}). These functions satisfy the boundary
condition
\begin{equation}\label{bcPhi}
\dot{\F}^\a(\t)\vert_{t=t_i}\ =\ 0
\end{equation}
as a consequence of the junction conditions (\ref{bcH}) and
(\ref{Phi}). They define an orthonormal basis with respect to the
scalar product:
\begin{equation}\label{scalprod}
\int_{-\pi}^\pi a(\t) \F^\a(\t)\F^\b(\t) dt\ =\ \delta^{\a\b}
\end{equation}
The explicit forms for these functions, when $\lambda = 0$, have
already been  obtained in
\cite{RSsecond} and \cite{GWphi} (See also \cite{Rizzo}).

The equations
(\ref{eqHA}) are simply the equations of motion for spin-$2$ fields
with mass $m_\a$ on a background which is an Einstein-space \cite{PF}
(for a 
detailed discussion of the flat case see also \cite{Veltman}). In
\cite{PF} it is shown that, when $m^2_\a >0$,
\begin{eqnarray}
&& \AH_{mk;a}^{~~~~~;a}-2\u{R}_{ambk}\AH^{ab}-m^2_\a \AH_{mk}\ =\ 0 \label{waveH} \\
&& \AH\ =\ \AH_{mk}^{~~;k}\ =\ 0\ \label{gaugeH}
\end{eqnarray}
and, therefore, we have five propagating degrees of freedom. Here,
$\u{R}_{ambk}$ is the Riemann tensor constructed from the metric
$\u{g}_{mk}$. On the other hand, when $m_\a=0$, the theory is
invariant under the gauge transformation
\begin{equation}\label{gaugetransf}
H_{mk}\ \to \ H_{mk} +\xi_{m;k}+\xi_{k;m}
\end{equation}
and describes a massless spin-2 field. In \cite{PF} it
was also demonstrated that the masses $m_\a$ of the states in the KK
tower have to satisfy
\begin{equation}\label{inequality}
\l +6m^2_\a\ \neq\ 0
\end{equation}

Using (\ref{Phi}) and
(\ref{eqPhi}), (\ref{adot}), (\ref{addot}), (\ref{bcPhi}) and
(\ref{scalprod}), we can rewrite the action (\ref{actionH}) in terms of the KK
states $\AH_{mk}$.  After some algebra, we obtain
\begin{equation}\label{actionKK}
S^{(2)}(d=0)\ =\ -2M^3 r_c \sum_\a \int d^4x \sqrt{-\underline{g}}
\int_{-\pi}^\pi dt\ {a\over 2}(\F^\a)^2\ L_{PF}(\AH_{mk};m^2_\a)
\end{equation}
where $L_{PF}$ is the generalization of the
Pauli-Fierz Lagrangian for a background which is an Einstein space
\cite{PF}, namely,
\begin{eqnarray}
L_{PF}(H,m^2)\ &=&\ {1\over 2} H_{ak;m}H^{ak;m}-H_{ak;m}H^{am;k}
+H^{mk}_{~~;k}H_{,m}-{1\over 2}H^{,m}H_{,m} \nonumber \\
&+& {m^2\over 2}\left( H_{mk}H^{mk}- H^2\right) + {\l\over 8}
\left( 2H_{mk}H^{mk}- H^2\right)\label{PFlag}
\end{eqnarray}
It is easy to check that the equations of motion  following from
(\ref{PFlag})  are precisely the same as (\ref{eqHA}).

\end{document}